\documentclass[aip, reprint]{revtex4-1}
\usepackage[english]{babel}

\usepackage{graphicx}
\usepackage{dcolumn}
\usepackage{bm}

\usepackage{hyperref}
\hypersetup{
 bookmarks=true,		
 unicode=false,			
 pdftoolbar=true,		
 pdffitwindow=false,		
 pdfstartview={FitH},		
 pdfcreator={pdflatex},		
 pdfnewwindow=true,		
 colorlinks=true,		
 linktoc=page,			
 linkcolor=blue,		
 citecolor=blue,		
 filecolor=blue,		
 urlcolor=blue			
}

\usepackage[utf8]{inputenc}
\usepackage[T1]{fontenc}
\usepackage{mathptmx}
\usepackage{etoolbox}
\usepackage{multirow}
\usepackage{siunitx}

\makeatletter
\def\@email#1#2{%
 \endgroup
 \patchcmd{\titleblock@produce}
  {\frontmatter@RRAPformat}
  {\frontmatter@RRAPformat{\produce@RRAP{*#1\href{mailto:#2}{#2}}}\frontmatter@RRAPformat}
  {}{}
}%
\makeatother

\def\NVm{NV$^-$}
\def\NV0{NV$^0$}
\def\Ns{N$_{\rm{s}}$}
\def\N0{N$_{\rm{s}}^0$}

\def\TA2{$^{\mathrm{3}}\!A_{\mathrm{2}}$}
\def\SA{$^{\mathrm{1}}\!A_{\mathrm{1}}$}
\def\SAp{$^{\mathrm{1}}\!A_{\mathrm{1}}'$}
\def\TE{$^{\mathrm{3}}\!E $}
\def\TEp{$^{\mathrm{3}}\!E' $}
\def\SE{$^{\mathrm{1}}\!E $}
\def\SEp{$^{\mathrm{1}}\!E' $}
\def\SEpp{$^{\mathrm{1}}\!E'' $}

\def\DA2{$^{\mathrm{2}}\!A_{\mathrm{2}}$}
\def\DTT{$\Delta$T/T}

\begin{document}

\title{Identifying high-energy electronic states of \NVm \ centers in diamond}

\author{Minh Tuan Luu}
\thanks{These authors contributed equally}
\affiliation{Max-Planck Institut f\"ur Polymerforschung, Ackermannweg 10, 55128 Mainz, Germany}

\author{Christopher Linder\"alv}
\thanks{These authors contributed equally}
\affiliation{University of Oslo, Department of Physics, Centre for Material Science and Nanotechnology, P.O. Box 1048, Blindern, Oslo N-0316, Norway}

\author{Zsolt Benedek}
\thanks{These authors contributed equally}
\affiliation{Department of Physics of Complex Systems, Eötvös Loránd University, Egyetem tér 1-3, H-1053 Budapest, Hungary}
\affiliation{MTA–ELTE Lend\"{u}let "Momentum" NewQubit Research Group, Pázmány Péter, Sétány 1/A, 1117 Budapest, Hungary}

\author{Ádám Ganyecz}
\affiliation{HUN-REN Wigner Research Centre for Physics, PO Box 49, H-1525, Budapest, Hungary}
\affiliation{MTA–ELTE Lend\"{u}let "Momentum" NewQubit Research Group, Pázmány Péter, Sétány 1/A, 1117 Budapest, Hungary}

\author{Gergely Barcza}
\affiliation{HUN-REN Wigner Research Centre for Physics, PO Box 49, H-1525, Budapest, Hungary}
\affiliation{MTA–ELTE Lend\"{u}let "Momentum" NewQubit Research Group, Pázmány Péter, Sétány 1/A, 1117 Budapest, Hungary}

\author{Viktor Ivády}
\affiliation{Department of Physics of Complex Systems, Eötvös Loránd University, Egyetem tér 1-3, H-1053 Budapest, Hungary}
\affiliation{MTA–ELTE Lend\"{u}let "Momentum" NewQubit Research Group, Pázmány Péter, Sétány 1/A, 1117 Budapest, Hungary}
\email{ivady.viktor@ttk.elte.hu}

\author{Ronald Ulbricht}%
 \email{ulbricht@mpip-mainz.mpg.de}
\affiliation{Max-Planck Institut f\"ur Polymerforschung, Ackermannweg 10, 55128 Mainz, Germany}%

\date{\today}

\begin{abstract}
The negatively charged nitrogen-vacancy center in diamond is a prototype photoluminescent point defect spin qubit with promising quantum technology applications, enabled by its efficient optical spin polarization and readout. Its low-lying electronic states and optical spin polarization cycle have been well characterized over decades, establishing it as a benchmark system for state-of-the-art computational methods in point defect research. While the optical cycle is well understood, a comprehensive energetic analysis of higher-lying states has received less attention until recently. In this joint experimental theoretical study, we identify and characterize five high-energy states beyond those involved in the optical cycle. Using transient absorption spectroscopy, we determine their transition energies and relative oscillator strengths. Additionally, we perform two independent numerical studies employing two state-of-the-art post-DFT methods to support the experimental findings and assign energy levels. These results enhance our understanding of the NV center’s energy spectrum and providing a broader reference for benchmarking high-level first-principles methods.
\end{abstract}

\maketitle

Nitrogen-vacancy (NV) centers in diamond \cite{DohertyNVreview} have emerged as a promising platform for quantum sensing \cite{DegenRMP2017,barry_sensitivity_2020,Aslam2023} and quantum information processing.\cite{wehner_quantum_2018} The negatively charged state (\NVm) is of particular interest due to its long spin coherence time at room temperature \cite{Balasubramanian:NatMat2009} and the ability to optically initialize and read out spin states, \cite{DohertyNVreview} allowing measurements of physical quantities such as electromagnetic fields,\cite{michl_robust_2019} temperature,\cite{Kucsko2013} and mechanical strain.\cite{ovartchaiyapong_dynamic_2014} Additionally, nearby nuclear spins enable the implementation of small-scale quantum computing circuits, while \NVm\ center implemented spin-photon interfaces and quantum repeaters hold promise for building quantum networks.

The spin-dependent optical excitation cycle, which enables optical initialization and readout, is governed by the low-lying electronically excited states of \NVm. These states, along with their optical and spin-orbit coupling, have been thoroughly studied, leading to a solid understanding of the spin polarization mechanism and the spin-dependent photoluminescence signal,\cite{Gali_2019} making this defect has become a benchmark system for state-of-the-art theoretical methods. \cite{Gali:PRL2009,Delaney2010,Ma2010,Choi2012,thiering_ab_2017,Thiering_2018,Bockstedte2018, Ma2020b,Bhandari2021,Sheng_2022,Barker2022,Haldar_2023,Ivanov_2023,Chen_2023, Li2024,benedek2024} Higher-lying excited states, which can play a role in photoionization and photoelectric readout of \NVm,\cite{gulka_room-temperature_2021} as well as extend the scope for theoretical benchmarking, have received much less attention.

Recently, we experimentally discovered a range of high-energy excited states using transient absorption (TA) spectroscopy.\cite{Luu24, Younesi2022} Our results also provided direct evidence of electron transfer from \NVm\ to nearby nitrogen defects, explaining photoluminescence quenching and its dependence on nitrogen concentration.\cite{Luu24} Theoretical predictions for high-energy states have been scarce,\cite{Zyubin2009,Maze2011,Doherty2011,Bhandari2021,Wirtitsch-2023} making their identification difficult. This is also because standard computational methods have difficulties in dealing with excited states, especially those beyond the first excited state that are potentially resonant with the conduction band (CB). The situation has improved with the development of new post-density functional theory (DFT) methods that can simultaneously account for the static correlation of the localized defect states and the screening effect of the embedding host electron states.

In this Letter, we further investigate the newly observed excited state by experimentally determining the relative transition strengths and conducting state-of-the-art numerical simulations using two independent methods to identify and label features of the high-energy spectra. 

\begin{figure*}[t]
    \centering
    \includegraphics[width=0.9\linewidth]{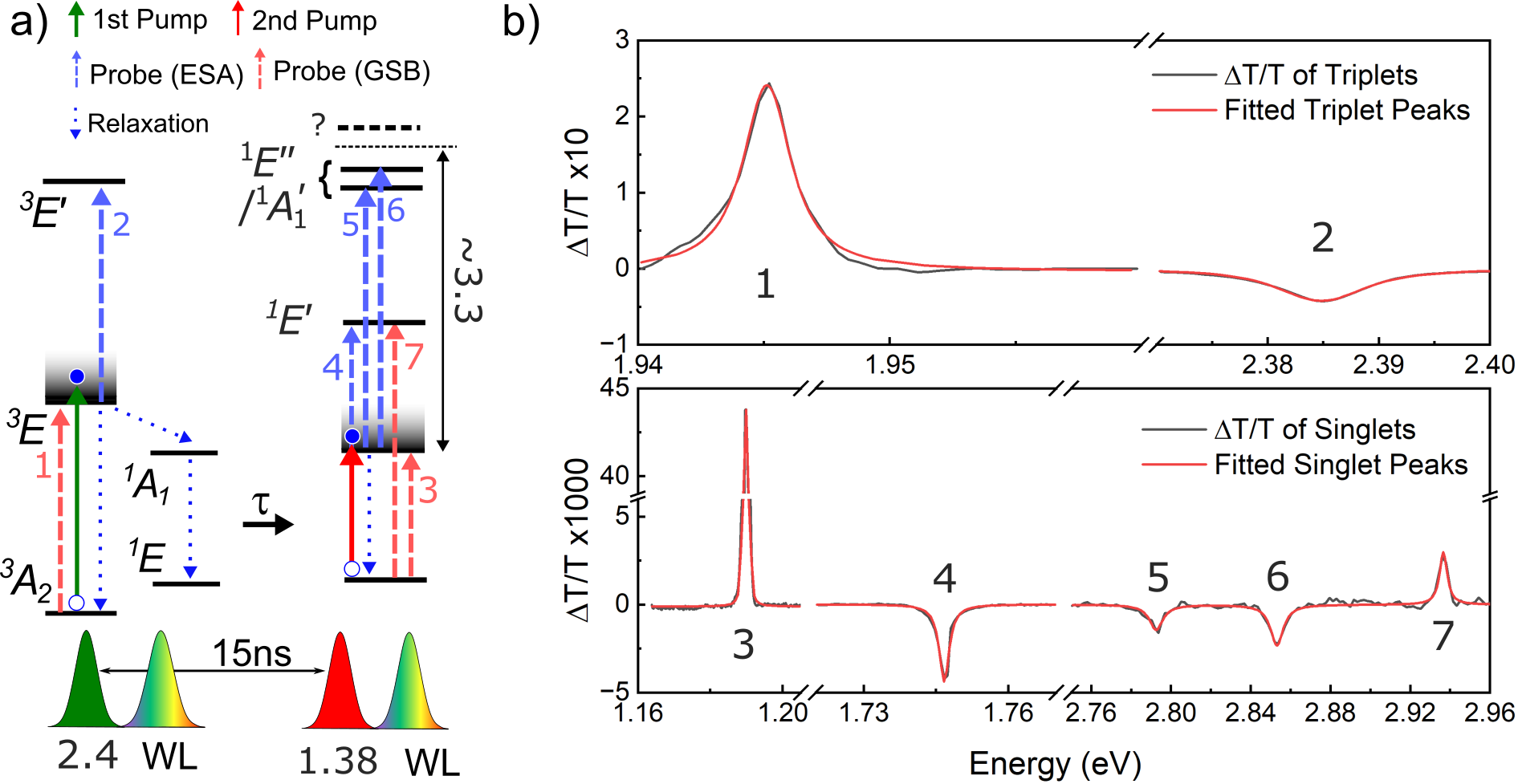}
    \caption{a) Schematic energy level structure and scheme of the TA measurements in triplet and singlet channels. Probe energy is limited to 3.3\,eV from \SA. b) Fitted curves of baseline-subtracted TA signal of transitions involving high energy states in \NVm. Top: triplet channel; Botton: singlet channel. Fit values are detailed in Table \ref{tab:allpeaks}. }
    \label{fig:singlet-expt}
\end{figure*}

A two-pump femtosecond TA setup was implemented to measure the excited electronic states of \NVm\ in both the triplet and singlet spin channels, shown schematically in Fig.~\ref{fig:singlet-expt}a. The first pump has a photon energy of 2.4\,eV (515\,nm), followed by a second pump at 1.38\,eV (900\,nm) arriving 15\,ns later. The white-light (WL) probe beam is generated in a 4\,mm c-cut sapphire crystal, with a spectrum covering either the 1.5 - 3.3\,eV range (2.4\,eV input, for higher excited states measurements) or the 1.1 - 2.3\,eV range (1.07\,eV input, for lower singlet states measurements). We probe the change in the transmission of the WL pulse (\DTT) under excitation, where \DTT\ > 0 indicates either ground state bleaching (GSB) or stimulated emission (SE) from the excited state, while \DTT\ < 0 signifies excited state absorption (ESA). To probe the triplet channel, the WL measures \DTT\ 5\,ps after the first pump (2.4\,eV). To measure the singlet channel, the system is allowed to relax after 2.4\,eV excitation for 15 ns, so that population has either decayed to \TA2\ or \SE. The second pump at 1.38\,eV then excites the \SE\ $\rightarrow$\ \SA\ transition, with the white-light probing \DTT\ 5\,ps later.

We used HPHT diamonds that contain about 2ppm of \NVm\ with a high concentration of single-substitutional nitrogen \Ns\ ($\sim$100\,ppm) to minimize the presence of \NV0. All measurements were performed at a temperature of 10\,K. Details of the experimental setup were previously published.\cite{Luu24}

To theoretically describe the observed high-energy states two approach was used: (i) complete active space self-consistent field (CASSCF) method~\cite{Siegbahn_1980,Roos_1980,Siegbahn_1981} followed by 2nd-order N-electron valence state perturbation theory (NEVPT2)~\cite{nevpt2,Angeli-2001a,Angeli2007} method on hydrogen-terminated nanodiamonds; (ii) constrained random phase and full-configurational interaction (cRPA+FCI) calculations \cite{Muechler_2022, Bockstedte2018} on supercells.

The first methodology is similar to that described in detail in our previous work.\cite{benedek2024} The used nanodiamonds, labeled as MODEL-I, MODEL-II, and MODEL-III, contain 34, 70, and 164 C atoms, and differ from our previous models\cite{benedek2024} by that they are expanded to reduce the steric clash among terminal H atoms. State-average (SA) CASSCF calculations, followed by NEVPT2, were performed with the 4 defect-localized band-gap orbitals and 6 electrons in the active space, with 5 triplet and 8 singlet eigenstates. We applied the cc-pVDZ~\cite{cc-pVDZ} basis set along with def2/J~\cite{Weigend2006} and cc-pVDZ/C~\cite{Weigend2002} auxiliary sets for density fitting approximations.\cite{RN29,RN56,RN192} The cluster geometries were optimized to the ground triplet ($^3A_2$) electronic state also with CASSCF. CASSCF-NEVPT2 calculations were carried out using ORCA 6.0.1.\cite{RN204,RN155,RN205} See SI, section S1.4 for sample input files. Here, results with MODEL-III are presented, further results with smaller models are also in the SI.

The second method is using DFT and constrained random phase approximation (cRPA) to parameterize a Hamiltonian that can be solved using full configuration interaction (FCI). This method was demonstrated for the low-energy states of the \NVm\ center recently.\cite{Muechler_2022}
The cRPA+FCI calculations \cite{Bockstedte2018} were performed as outlined in Ref. \citenum{Muechler_2022} using VASP \cite{VASP, VASP2} and an in-house implementation of configuration interaction.
Here, the lattice parameter was 3.57~{\AA} and a $3\times 3\times 3$ repetition of the unit cell was used for modeling the defect, consisting of 216 atoms. The exchange-correlation functional used was HSE06. \cite{HSE03,HSE06} The Brillouin zone was sampled at the $\Gamma$-point only and the plane wave basis was truncated at 520~eV. In the cRPA\cite{Aryasetiawan04} calculation, 4 localized (spin-paired) states were used and empty states with energies up to 36~eV above the CB were included. The one-body matrix elements were obtained from orbital energy levels with Hartree double counting correction.\cite{Muechler_2022}

Figure \ref{fig:singlet-expt} shows the energy scheme of the electronic states and optical transitions in both the singlet and triplet channels. Exciting electrons from \TA2\, to \TE\, using a 2.4\,eV pump revealed an ESA feature at 2.38\,eV (520\,nm) whose relaxation dynamics matched that of the SE signal from the \TE\ state, indicating that the ESA signal observed is a spin-triplet state above \TE\ and was attributed to the \TEp\, state. This \TEp\ state also appeared in photoluminescence excitation (PLE) measurements\cite{beha12} as a reduction of the \NVm\ PL signal and an increase in \NV0 signal, hinting towards a charge conversion process as the \TEp\ got excited. 

For the singlet system, the TA spectrum reveals several absorption peaks at 1.75\,eV, 2.79\,eV and 2.85\,eV that share the relaxation lifetime of 130\,ps of the \SA\ state.\cite{ulbricht18,Luu24} The spectrum also shows the \SE$\rightarrow$\SEp\ transition at 2.94\,eV as GSB. Although 1.75\,eV has been predicted before\cite{gali08,Larsson:PRB2008,Doherty2011} and experimentally verified along with a GSB replica at 2.94\,eV, \cite{Luu24} the higher energy singlet states have neither been theoretically considered nor concretely observed experimentally. Due to the rapidly diminishing probe sensitivity towards the UV region, any potential transition located 3.3\,eV above \SA\ can not be identified. 

All ZPL peaks were fitted with Lorentzian functions, giving their spectral positions, linewidth, and relative intensity. The intensity of the peaks in the triplet or singlet channel was normalized to the transition between the two lowest states in each system (\TA2$\rightarrow$\TE\, for triplets and \SE$\rightarrow$\SA\ for singlets). The results of the peak fitting are summarized in Table~\ref{tab:allpeaks} alongside the values from theoretical models. Due to the low intensity of the higher singlet state peaks, the phonon sideband of such transitions could not be observed. The intensity of the peaks at the ZPLs comprises both the GSB and SE signals. However, this was only relevant in the \TA2$\rightarrow$\TE\, transition since the SE signal for this transition was comparable to the GSB signal. For other transitions, the SE signals were extremely low compared to the GSB signal or simply not detectable and thus excluded in the oscillator strength values.

\begin{figure}[h]
    \centering
    \includegraphics[width=0.7\linewidth]{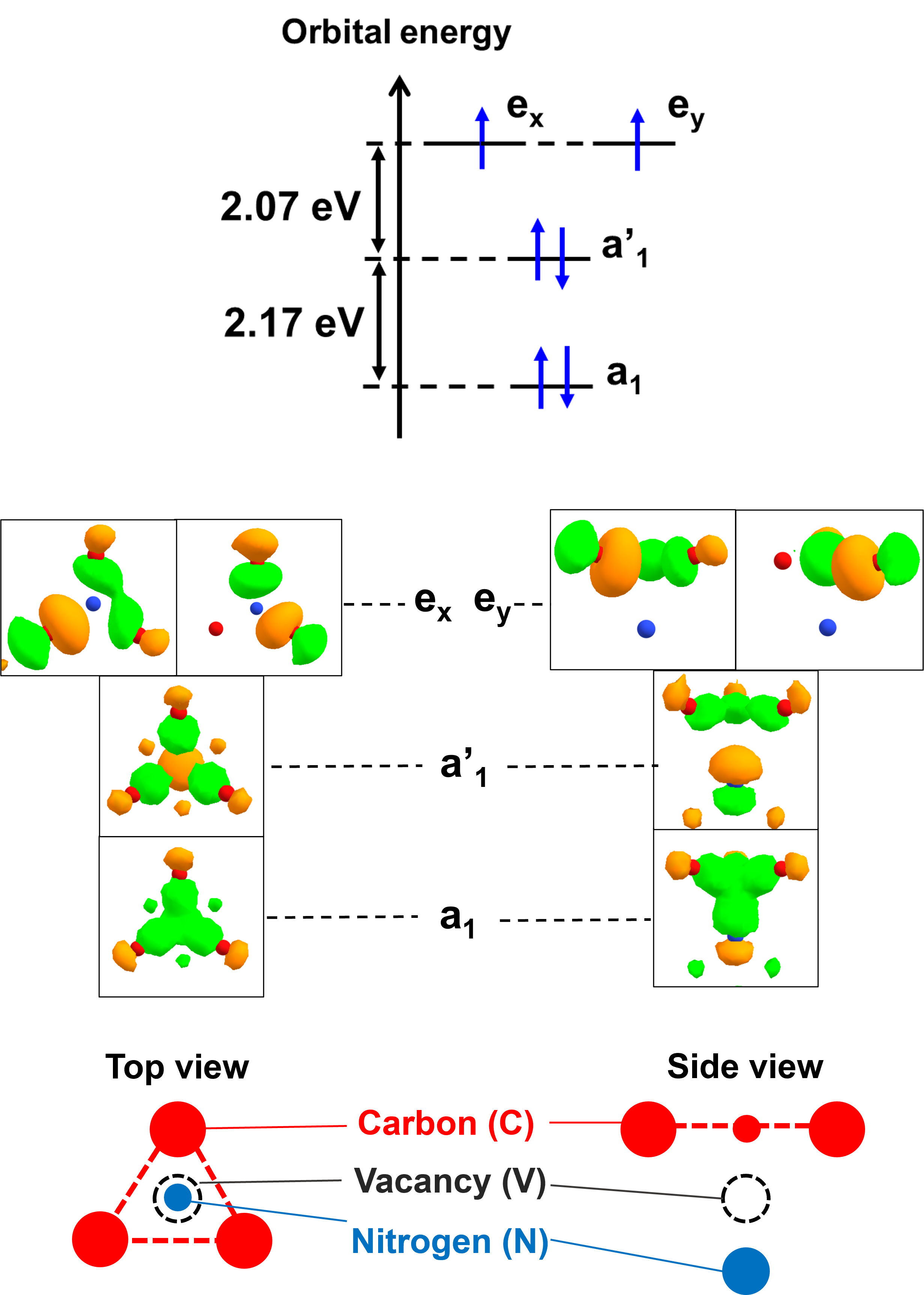}
    \caption{Top: Relative energy levels of the defect-localized orbitals of \NVm, as obtained upon wannierization, i.e. projection of localized spin paired HSE06 orbitals of the supercell model on sp3 states of the nearest neighbour atoms. 
    Bottom: surface plot of the localized band-gap orbitals in the vicinity of the vacancy. Green and orange surfaces indicate positive and negative wavefunction sign, respectively. (Note that CASSCF on cluster models provides similar defect orbital diagram.)
    }
    \label{fig:theory_orbitals}
\end{figure}

\begin{figure}[h]
    \centering
    \includegraphics[page=3,width=1.00\linewidth]{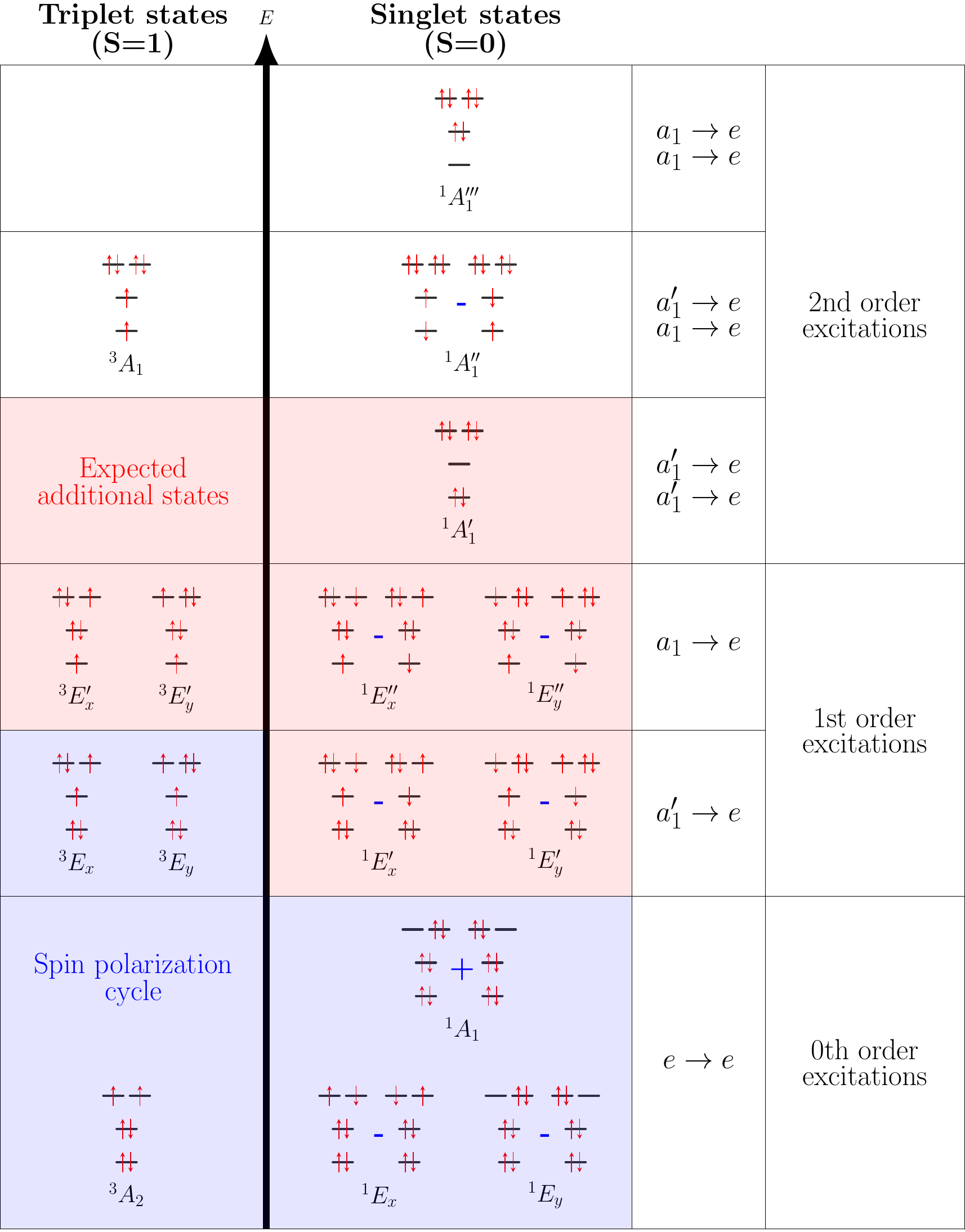}
    \caption{Expected electron configurations for the distribution of 6 electrons on the 4 band-gap orbitals according to group theory. The configurations are ordered - from bottom to top - according to the expected energy levels based on the orbital energy diagram. }
    \label{fig:theory_configurations}
\end{figure}

We start the theoretical considerations by discussing the possible nature of high-energy states using group theory, based on the orbital energy diagram of the NV$^-$ center (see Fig.~\ref{fig:theory_orbitals}). There are four defect-localized orbitals in the vicinity of the band gap, formed by the superposition of four dangling atomic $2p$ orbitals around the vacancy. 
Using notations corresponding to the $C_{3v}$ symmetry, we have two half-occupied degenerate $e$ orbitals and two occupied $a_1$ orbitals ($a_1$ and $a'_1$) in the ground state $\sim$2 eV apart from each other. 
If we assume that all band-gap electronic states of NV$^-$ are formed by excitations within the latter four orbitals, 5 triplet and 10 singlet excited states are conceivable in addition to $^3A_2$, see Fig.~\ref{fig:theory_configurations}. 

In the bottom row, we show the occupation patterns where the $e$ and $a_1$ orbitals are semi- and doubly occupied, respectively. These can be understood as "zeroth-order" excitations, which are expected to have the lowest energy level among all states. These states are $^3A_2$, $^1E$, and $^1A_1$.

For first-order excitations only $a'_1\rightarrow e$ and $a_1\rightarrow e$ electron jumps are feasible and the orbital energy gap in the former case is approximately twice smaller. The $a'_1\rightarrow e$ process leads to $^3E$ and $^1E'$ excited states.
Among these, $^3E$ is responsible for the 1.945 eV emission line of NV$^-$~\cite{Gali-th2009} and is known to be involved in its spin polarization cycle~\cite{Gali_2019} ($^3A_2 \rightarrow {}^1A_1 \rightarrow {}^1E \rightarrow {}^3A_2$; highlighted by blue background in Fig.~\ref{fig:theory_configurations}). However, the existence of the singlet analog $^1E'$ has only been theoretically predicted to date.\cite{Gali_2019,Wirtitsch-2023} Nevertheless, as the group theory indicates that the energy level of $^1E'$ is commensurable with that of $^3E$, it is reasonable to assume that $^1E'$ is present in the TA spectrum. 
The $a_1\rightarrow e$ transition produces another set of $E$ states, namely $^3E'$ and $^1E''$.

We got two triplet and three singlet excited states; however, the three ESA peaks in the singlet TA spectrum indicate that at least one doubly excited singlet state should also be accessible by absorption. Considering the possible second-order excitations, see the top of Fig.~\ref{fig:theory_configurations}, energetically the double $a'_1 \rightarrow e$ process is the most favorable. According to the Kohn-Sham orbital picture, the energy demand of a double $a'_1 \rightarrow e$ excitation is similar to a single $a_1 \rightarrow e$ jump. 

Altogether, chemical intuition and group theory strongly suggest that the additional states observed in TA spectroscopy are $^3E''$, $^1E'$, $^1A'_1$ and $^1E''$. These states are highlighted with red background in Fig.~\ref{fig:theory_configurations}. Furthermore, given that only four defect-localized orbitals are present, any other possible distribution of electrons results in double excitations of significantly higher energy.

Now, we test the qualitative predictions of group theory by ab initio calculations. Our previously reported initial CASSCF-NEVPT2 results~\cite{benedek2024} indicated that the newly suggested states of $^3E''$, $^1E'$, $^1A'_1$, and $^1E''$ are located in the energy range of 3.3-4.6 eV relative to the ground triplet state ($^3A_2$), which correspond well with our present measurement. Herein, we further investigate our hypothesis on $^3E''$, $^1E'$, $^1A'_1$, and $^1E''$ states by (i) a comparative analysis of CASSCF-NEVPT2 and CRPA+FCI, (ii) comparing theoretical oscillator strengths from CASSCF-NEVPT2 to the observed relative peak sizes. 
 
As both theoretical approach takes state mixing effects into account, it is worth studying the composition of the eigenstates based on the group theoretically pure configurations of Fig.~\ref{fig:theory_configurations}, see Tabs. S3-S4 of SI.
Strikingly, state mixing effects are observed in all configurations of the singlet channel, in sharp contrast to the triplets where all states were found to be dominated by one configuration.
For instance, we observe the previously reported~\cite{Thiering_2018} presence of $^1E'$ in $^1E$, which enables the intersystem crossing to $^3A_2$, as well as the previously reported~\cite{Bhandari2021,li2024excitedstatedynamicsopticallydetected} significant contribution of $^1A'_1$ to the multireference wavefunction of $^1A_1$. We emphasize here that the multireference character of singlets makes the single-determinant methods, such as DFT, highly unreliable for calculating energy differences. 

\begin{table*}[ht]
\setlength\extrarowheight{3pt}
\setlength{\tabcolsep}{0.8em}
    \centering 
    \begin{tabular}{cl|ccc|ccc|c}
    \hline
    \hline
         \multirow{2}{*}{Peak}  & \multirow{2}{*}{Transition} & \multicolumn{3}{c|}{Energy (eV)}   &  \multicolumn{3}{c|}{Oscillator strength}                & DW (\%) \\
        &            & exp.   & theory 1      &  theory 2     & exp.   & theory 1     & theory 2  &  exp. \\             
    \hline
    \hline
        1 & \TA2\ $\rightarrow$\ \TE        & 1.945 [1.4] & 2.11 & 2.20 & 1* & 1 & - &3.2 \cite{alkauskas14,kehayias13}\\
        2 & \TE\  $\rightarrow$\ \TEp       & 2.385 [7.7] & 2.22 & 2.35 & 0.49 & 0.54 & - &8.1 \cite{Luu24} \\
    \hline
        3 & \SE\  $\rightarrow$\ \SA        & 1.19 [1.4] & 1.02 & 0.88 & 1 & 1 & -  &41 \cite{kehayias13}\\
        4 & \SA\  $\rightarrow$\ \SEp       & 1.75 [3.3] & 1.71 & 1.82 & 0.22 & 0.04 & - &42\cite{Luu24} \\
        5 & \SA\  $\rightarrow$\ \SEpp/\SAp & 2.79 [6.9] & 2.89 & 3.61 & 0.01 & 0.10 & - &-\\
        6 & \SA\  $\rightarrow$\ \SEpp/\SAp & 2.85 [8.1] & 2.35 & 3.77 & 0.02 & 1.14 & - &-\\
        7 & \SE\  $\rightarrow$\ \SEp       & 2.94 [4.7] & 2.73 & 2.70 & 0.01 & 0.10 & - &-\\
        \hline
        \hline
    \end{tabular}
    \caption{Summary of observed transition peak  properties (Fig.~\ref{fig:singlet-expt}b)  compared to CASSCF-NEVPT2 (theory 1) and CRPA+FCI (theory 2) results.  Transition energies are in eV, full-width-half-maximum (FWHM)  of the experimental peaks (in square brackets) are measured in meV. 
    Oscillator strengths of the singlet and the triplet systems are normalized to the corresponding ZPL cross-sections.  
    *: Intensity halved assuming GSB and SE give equal contributions. 
   }
    \label{tab:allpeaks}
\end{table*}

\begin{figure}[h]
    \centering 
    \includegraphics[page=1,width=1.00\linewidth]{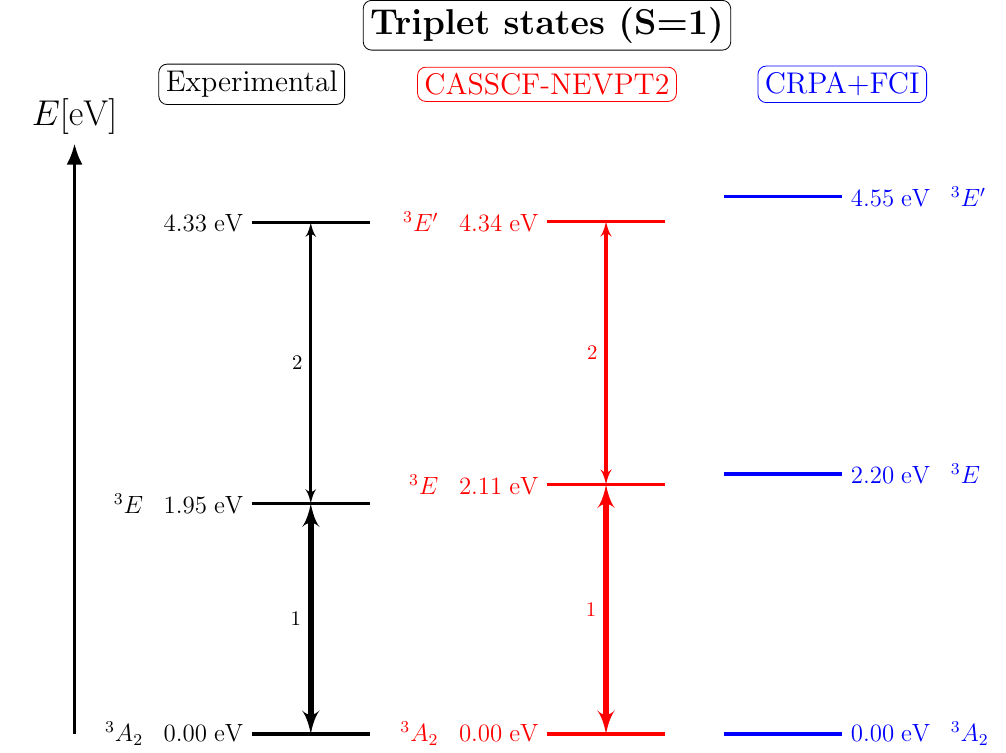}
    \includegraphics[page=2,width=1.00\linewidth]{en_diag.pdf}
    \caption{Comparison of experimental (black) and theoretical (red, blue) energy levels of the electronic states of NV$^-$ in triplet (top graph) and singlet (bottom graph) spin state. Experimental levels were calculated from observed ZPL wavelengths, while theroetical values refer to vertical gaps, calculated at CASSCF-NEVPT2 and CI + CRPA level, respectively. (See SI, section S3.1 for a discussion of comparability of the three methods.) The thickness of arrows is proportional to the oscillator strength. 
    }
    \label{fig:theory_comparison}
\end{figure}

Next, we turn our focus on excitation energies as summarized in Tab.~\ref{tab:allpeaks}, left and visualized in Fig.~\ref{fig:theory_comparison}; see also Tabs. S1-2 of SI for a detailed data analysis.  
With CASSCF-NEVPT2 we obtain triplet electronic energy levels of 2.11 eV ($^3E$) and 4.34 eV ($^3E''$), relative to the ground state, which corresponds to absorption peaks of 2.11 eV ($^3A_2 \rightarrow {}^3E$) and 2.22 eV ($^3E' \rightarrow {}^3E''$), respectively. The results agree well with the experimental peak positions of 1.95 eV and 2.39 eV, respectively. A similarly successful reproduction of the measured energy levels was achieved in the singlet channel, where the excited states are located, compared to the lowest-lying state of $^1E$, at 1.02 eV ($^1A_1$), 2.73 eV ($^1E'$), and 3.91 eV ($^1A'_1$). 
The only outlier is the $^1E''$ state with 3.37 eV. Thus, within the measurement limit of <3.3\,eV, we predict absorption at 1.02 eV ($^1E \rightarrow {}^1A_1$) and 2.73 eV ($^1E \rightarrow {}^1E'$) from the ground singlet state, while 1.71 eV ($^1A_1 \rightarrow {}^1E'$), 2.89 eV ($^1A_1 \rightarrow {}^1A'_1$) and 2.35 eV ($^1A_1 \rightarrow {}^1E''$) from the first excited singlet. 

With cRPA+FCI the excitation energies in the triplet channel are 2.20~eV ($^3A_2\rightarrow {}^3E$) and 4.55~eV $(^3A_2\rightarrow {}^3E')$. The $^3E\rightarrow {}^3E'$ transition energy is at 2.35~eV. These energies agree very well with both experimental measurements and energies obtained with CASSCF-NEVPT2. For the singlet channel, the excitation energy to the first two states ($^1A_1$ and $^1E'$) are at 0.88~eV and 2.70~eV, respectively. The next state is at 4.49~eV above the lowest singlet state ($^1E$). This state is the $^1A_1'$ state, with the $^1E''$ state located 0.15~eV above the $^1A_1'$ state. This is in contrast to the position, ordering and splitting obtained with CASSCF-NEVPT2. The energy splitting of 0.15~eV between $^1A_1'$ and $^1E''$ is reminiscent of the energy splitting between the experimentally observed states at 3.98~eV and 4.04~eV, but the excitation energy is about 0.5~eV larger than experimentally observed. Furthermore, the $^1E''$ state described with cRPA+FCI is more than 1~eV above the value obtained with CASSCF-NEVPT2. The cRPA+FCI method involves certain approximations that may influence the positioning of energy levels. First, the electronic screening is in this case computed with the static dielectric function. Furthermore, in the present calculations, there are some double counting errors in the one-body matrix elements (Kohn-Sham energy levels) due to the exchange-correlation energy of the active space, which should be fully included in the two-body matrix elements, but which is neglected in this case. The present cRPA  calculation, with the considered minimal basis set is qualitatively in agreement with experiements.

Altogether, CASSCF-NEVPT2 predicts the newly identified TA peak positions with less than 0.2 eV deviation from the measurement, with the exception of $^1E''$, which differs by 0.5 eV. This finding is strong evidence for the correctness of our assignment of higher-energy states shown in Fig.~\ref{fig:theory_configurations}.

Apart from the transition energies, the transition oscillator strength ($f_{\rm osc}$), i.e. absorption cross-section, could be used to determine the electronic states. (See SI, section S3.2 for details.) The calculated oscillator strengths, compared to the experimental integrated peak areas, are provided in Table~\ref{tab:allpeaks}. The values are normalized, the same way as the experimental ones. In our comparison, we expect only an order-of-magnitude estimate from theory, given the approximations involved in our methods and the fact that the experimental values of $f_{\rm osc} $ are based only on ZPL peaks, even though the Debye-Waller factor could significantly vary.

In the triplet channel, we find $f_{\rm osc} = 0.54$ for $^3E \rightarrow {}^3E'$ at CASSCF-NEVPT2 level, which is in good agreement with the experimental value of 0.49. In the singlet channel, the peak intensities show more variation. According to our theoretical calculations, after the $^1E \rightarrow {}^1A_1$ transition feature of the TA spectrum is the $^1A_1 \rightarrow {}^1E'$ transition ($f_{osc}$ = 0.04), followed by the $^1E \rightarrow {}^1E'$ ($f_{\rm osc}$ = 0.10) and the $^1A_1 \rightarrow {}^1A_1'$ ($f_{\rm osc}$ = 0.10) transitions. The deviations of the oscillator strengths from the experimental values, 0.22, 0.01, and 0.01, respectively, are tolerable in these cases.

The $^1A_1 \rightarrow {}^1E''$ transition is an exceptional case, where the obtained $f_{\rm osc}$ value of 1.14 is almost two orders of magnitude higher than the experimental observations. 
Additionally, the energy of the corresponding transition is in disagreement with the experimental results. This casts some doubt on the identity of one of the peaks at 2.79-2.85 eV, as such a deviation cannot likely be explained by DW-factor variations or by any factor neglected by our computational protocol (e.g. geometry relaxation effects). 
The potential causes of this discrepancy are discussed in the SI, section S3.3. For the cRPA+FCI calculations, the gauge for the position operator in the Wannier basis does not coincide with the center of charge density and hence the optical matrix elements and consequently the oscillator strengths are not evaluated with this level of theory.

In conclusion, our joint experimental and theoretical study elaborates on the high-energy spectra of the NV center in diamond. Our results, obtained through both TA spectroscopy and advanced computational methods, provide a comprehensive characterization of five high-energy states beyond the commonly studied optical cycle. By combining experimental transition energies and oscillator strengths with theoretical insights from CASSCF-NEVPT2 and cRPA+FCI methods, we not only provide the identification of these states but also deepen our understanding of the NV center’s electronic structure. 

These findings can have important implications for future applications of NV centers in quantum technologies, particularly in the development of more efficient spin-to-charge conversion readout mechanisms. Additionally, the results we obtained offer a valuable framework for benchmarking high-level first-principles methods in the study of point defects.

\section*{Acknowledgments}

This research was supported by the National Research, Development, and Innovation Office of Hungary within the Quantum Information National Laboratory of Hungary (Grant No. 2022-2.1.1-NL-2022-00004) and grants FK 135496 and FK 145395. Z.B.\ also acknowledges the financial support of the János Bolyai Research Fellowship of the Hungarian Academy of Sciences. We acknowledge the support of the European Union under Horizon Europe for the QRC-4-ESP project (Grant Agreement 101129663), the QUEST project (Grant Agreement 101156088) and funding by the Max-Planck Society.
C.~L. acknowledges support provided by the Research Council of Norway and the University of Oslo through the research project QuTe (no. 325573, FriPro ToppForsk-program).

The computations were enabled by resources provided by the National Academic Infrastructure for Supercomputing in Sweden (NAISS) and the Swedish National Infrastructure for Computing (SNIC) at NSC, partially funded by the Swedish Research Council through grant agreement no. 2022-06725 and no. 2018-05973. We also acknowledge KIF\"U for awarding us computational resources at the Komondor supercomputer in Hungary. In addition, computations were performed on resources provided by UNINETT Sigma2 --- the National Infrastructure for High Performance Computing and Data Storage in Norway.

\end{document}


\maketitle
\date{}


\tableofcontents
\newpage

\section{Computational details of CASSCF-NEVPT2 calculations}

\subsection{Model size dependence of excitation energies and oscillator strengths}

\begin{table*}[ht]
\setlength\extrarowheight{3pt}
\setlength{\tabcolsep}{0.8em}
    \centering
    \resizebox{\textwidth}{!}{%
    \begin{tabular}{c|cccc|cccc|cccc}
    \hline
    \hline
         State/  & \multicolumn{4}{|c|}{MODEL-I } & \multicolumn{4}{|c|}{MODEL-II } & \multicolumn{4}{|c}{MODEL-III } \\
         \cline{2-13}
       Transition &  $E_0$ & $\Delta E$ & $E_{\rm tot}$  & $f_{\rm osc} $ &  $E_0$ & $\Delta E$ & $E_{\rm tot}$ & $f_{\rm osc}$  &  $E_0$ & $\Delta E$ & $E_{\rm tot}$ & $f_{\rm osc}$ \\             
    \hline
    \hline
      $^3A_2$ &0.00& 0.00&0.00&-&0.00& 0.00&0.00&-&0.00& 0.00&0.00&- \\
      $^3E$   &2.71&-0.70&2.01&-&3.08&-0.88&2.20&-&3.11&-1.00&2.11&- \\
      $^3E'$  &6.55&-1.91&4.65&-&6.91&-2.24&4.67&-&6.83&-2.50&4.34&- \\
        \hline
      $^3A_2 \rightarrow {}^3E$ &2.71&-0.70&2.01&0.20 (1.00)&3.08&-0.88&2.20&0.19  (1.00)&3.11&-1.00&2.11&0.20 (1.00) \\
      $^3E \rightarrow {}^3E'$  &3.84&-1.21&2.64&0.16 (0.82)&3.83&-1.36&2.47&0.16 (0.82)&3.72&-1.50&2.22&0.11 (0.54) \\
    \hline
    \hline
    \end{tabular}}
    \caption{Model size dependence of energy levels of excited states, transition energies and oscillator strengths in the triplet sector. All presented values were computed at CASSCF(6e,4o)-NEVPT2/cc-pVDZ level, with state-averaging over the 5 lowest-lying roots. Energy values (CASSCF energy ($E_0$), NEVPT2 correction ($\Delta E$) and total CASSCF-NEVPT2 energy ($E_{\rm tot}$), respectively) are given in eV unit. Oscillator strength ($f_{\rm osc}$) is dimensionless, parenthesized values are normalized to the $f_{\rm osc}$ of the optical transition of the spin polarization cycle.}
    \label{tab:SM-levels}
\end{table*}

\begin{table*}[ht]
\setlength\extrarowheight{3pt}
\setlength{\tabcolsep}{0.8em}
    \centering
    \resizebox{\textwidth}{!}{%
    \begin{tabular}{c|cccc|cccc|cccc}
    \hline
    \hline
         State/  & \multicolumn{4}{|c|}{MODEL-I } & \multicolumn{4}{|c|}{MODEL-II } & \multicolumn{4}{|c}{MODEL-III } \\
         \cline{2-13}
       Transition &  $E_0$ & $\Delta E$ & $E_{\rm tot}$  & $f_{\rm osc} $ &  $E_0$ & $\Delta E$ & $E_{\rm tot}$ & $f_{\rm osc}$  &  $E_0$ & $\Delta E$ & $E_{\rm tot}$ & $f_{\rm osc}$ \\            
    \hline
    \hline
      $^1E$ &0.81&-0.27&0.54&-&0.92&-0.35&0.57&-&0.92&-0.37&0.55&- \\
      $^1A_1$ &2.78&-1.12&1.66&-&3.11&-1.42&1.70&-&3.10&-1.53&1.57&- \\
      $^1E'$ &5.64&-2.11&3.54&-&6.01&-2.43&3.58&-&5.96&-2.69&3.28&- \\
      $^1A'_1$ &8.14&-3.33&4.81&-&8.87&-4.01&4.86&-&8.83&-4.37&4.46&- \\
      $^1E''$ &9.63&-4.69&4.94&-&10.21&-5.72&4.49&-&10.04&-6.13&3.92&- \\
        \hline
      $^1E \rightarrow {}^1A_1$ &1.97&-0.86&1.12&0.12 (1.00)&2.19&-1.06&1.13&0.10 (1.00)&2.18&-1.16&1.02&0.10 (1.00) \\
      $^1A_1 \rightarrow {}^3E'$ &2.87&-0.99&1.88&0.01 (0.04)&2.90&-1.02&1.88&0.01 (0.03)&2.86&-1.15&1.71&0.01 (0.04) \\
      $^1A_1 \rightarrow {}^1A'_1$ &5.36&-2.21&3.15&0.02 (0.14)&5.76&-2.60&3.16&0.01 (0.11)&5.73&-2.84&2.89&0.01 (0.10) \\
      $^1A_1 \rightarrow {}^1E''$ &6.85&-3.57&3.28&0.15 (1.22)&7.10&-4.30&2.80&0.13 (1.22)&6.95&-4.60&2.35&0.11 (1.14) \\
      $^1E \rightarrow {}^1E'$ &4.84&-1.84&2.99&0.02 (0.13)&5.09&-2.08&3.01&0.01 (0.14) &5.04&-2.31&2.73&0.01 (0.10) \\
    \hline
    \hline
    \end{tabular}}
    \caption{Similar to Tab.~\ref{tab:SM-levels} but for the singlet sector with state-averaging over 8 roots.}
    \label{tab:singlet_spectrum}
\end{table*}

\newpage

\subsection{Composition of CASSCF eigenstates (state mixing effects)}
\label{si:sect:casscf}
Table~\ref{tab:state_mixing} lists the composition of CASSCF eigenstates on the basis of group theoretically pure configurations. We note that the analysis is based on the multi-determinant picture on top of state-averaged CASSCF orbitals. As the optimal orbitals of individual states might vary, and can be significantly different from SA orbitals, the multi-state description presented in the Table does not always indicate actual multireference character (state mixing). For example,  the simultaneous presence of $^1E'$ and $^1E''$ (which differ in the occupation of $a_1$ and $a'_1$ orbitals of the same irreducible representation) likely derives from the unoptimal shape of state-averaged $a_1$ and $a'_1$ molecular orbitals, which could be resolved by state-specific orbital optimization. The same holds true for $^1A'_1$, $^1A''_1$ and  $^1A'''_1$ states.   

\begin{table*}[h]
\setlength\extrarowheight{3pt}
\setlength{\tabcolsep}{0.8em}
    \centering
    \begin{tabular}{c|c|l}
    \hline
    \hline
State \# & Group theory prediction & Composition of CASSCF eigenstate \\
    \hline
    \hline
1st triplet & $^3A_2$ & 100\% $^3A_2$ \\
2nd/3rd triplet & $^3E$ & \phantom{1}98\% $^3E\phantom{'}$ \\
4th/5th triplet & $^3E'$ & \phantom{1}98\% $^3E'\phantom{'}$, \phantom{1}1\% $^3E$\\
\hline
1st/2nd singlet & $^1E$ & \phantom{1}82\% $^1E\phantom{_1}$, 17\% $^1E'\phantom{'}$, \phantom{1}1\% $^1E''$ \\
3rd singlet & $^1A_1$ & \phantom{1}77\% $^1A_1$, 19\% $^1A'_1\phantom{'}$, \phantom{1}4\% $^1A''_1$  \\
4th/5th singlet & $^1E'$ & \phantom{1}46\% $^1E'\phantom{'}$, 49\% $^1E''$, \phantom{1}5\% $^1E$  \\
6th singlet & $^1A'_1$ & \phantom{1}74\% $^1A'_1$, 13\% $^1A_1\phantom{'}$, \phantom{1}3\% $^1A''_1$, 10\% $^1A'''_1$   \\
7th/8th singlet & $^1E''$ & \phantom{1}50\% $^1E''$, 36\% $^1E'\phantom{'}$, 14 \% $^1E$  \\
    \hline
    \hline   
    \end{tabular}
    \caption{Quantification of state mixing effects: composition of CASSCF eigenstates on the basis of gruop theoretically pure electron configurations. Only the leading configurations (weight $>$ 1\%) of each eigenstate are provided.}
    \label{tab:state_mixing}
\end{table*}

\subsection{Results using extended (8-orbital) active space}

To investigate the possible role of defect-localized virtual orbitals in forming additional band-gap states (i.e. apart from those shown in Fig. \textcolor{red}{2}), we ran follow-up CASSCF-NEVPT2 calculations with an extended active space (6 electrons on 8 orbitals). In particular, we added a the dangling $sp^3$ atomic orbitals of the inner C and N atoms from the 3rd electronic shell, which were identified upon the Pipek-Mezey localization of the virtual space. The state-averaged CASSCF orbital optimization for 5 triplet ($^3A_2, ^3E, ^3E'$) and 8 singlet ($^1E, ^1A_1, ^1E', ^1A'_1, ^1E''$) roots resulted in an orbital picture where the newly formed active orbitals (i.e. the 3rd-shell analogues of $e_{x/y}$, $a_1$ and $a'_1$) are located 16.7-25.2 eV above $e$, making the excitation of electrons to the 3rd shell highly unlikely. Nevertheless, we attempted to increase the number of roots to reveal any additional band-gap electronic states. The CASSCF orbital optimization, however, tends to rotate the 4 higher-energy orbitals out of the active space, and form unphysical orbitals that cover the edges of the cluster. This phenomenon was observed even for MODEL-III, and even using large level shifts.

\subsection{Sample input files (ORCA 6.0.1.)}

\subsubsection{Generation of initial (Kohn-Sham) orbitals}

! ROKS RI PBE cc-pVDZ def2/J \\

*xyz -1 3 [Geometry guess (xyz format)] * 

\subsubsection{Geometry optimization (SS-CASSCF)}

! cc-pVDZ moread def2/J opt \\
\%moinp ''KS\_orbitals.gbw'' \\
\%casscf \\
nel 6 \\
norb 4 \\
mult 3  \\
nroots 1 \\
end \\
\%geom \\
Constraints \{C \emph{number\_of\_fixed\_atoms} C\} end \quad \quad \quad  \#Fixed atoms: hydrogens and carbons of the outermost shell\\
end \\

* xyz -1 3 [Geometry guess (xyz format)] * \\ 

\subsubsection{Single-point energy calculation with state-averaging over all electronic states (SA-CASSCF-NEVPT2)}

! rijcosx cc-pVDZ def2/J moread cc-pVDZ/C \\

\%moinp ''CASSCF\_orbitals\_of\_optimized\_geometry.gbw'' \\
\%casscf \\
nel 6 \\
norb 4 \\
mult 3,1 \\
nroots 5,8 \\
bweight 5,8 \\
printwf det \\
actorbs canonorbs \\
ptmethod sc\_nevpt2 \\
dotrans all \\
nevpt \\
d3tpre 1e-14 \\
d4tpre 1e-14 \\
end \\
end \\

* xyz -1 3 [Optimized geometry (xyz format)] * \\

\subsection{Structure of cluster models used for WFT calculations}

The model was built by taking the experimental C-C bond length of diamond (1.5444 \AA{}), made of 35 ,71 or 165 C atoms and capped with H-s at the edges with 1.09\AA{} bond length. Then, a C was taken out and another one swapped to N (resulting in molecules of C$_{34}$NH$_{36}^-$, C$_{69}$NH$_{60}^-$ and C$_{164}$NH$_{100}^-$ formulas), and optimized at CASSCF(6e,4o)/cc-pVDZ level at $^3A_2$ electronic state. The outer H and C atoms were fixed, in order to simulate the solid-phase diamond environment.

The geometries of the optimized models in xyz format are provided below. Coordinats are given in \AA{} unit.\\

MODEL-I\\
\\
  N           0.99181339224083      0.99181088275911      0.99181283883722 \\
  C          -0.96114988905843     -0.96116332852481      0.94835665157548 \\
  C          -0.96127116291968      0.94850928699562     -0.96128243233846 \\
  C           0.94833827441120     -0.96116206993323     -0.96113504976517\\
  C           0.00000014763818      1.78334988193848      1.78334986168052\\
  C           1.78334986794356      1.78334986609157      0.00000015476413\\
  C           1.78334979585378      0.00000011837242      1.78334981066380\\
  C          -0.00000001221921     -1.78335001310501      1.78335001768926\\
  C          -1.78334998533266     -1.78334999250587     -0.00000006554514\\
  C          -1.78335003508102     -0.00000000662231      1.78335003061237\\
  C          -1.78335002468338      1.78335003076441      0.00000000706105\\
  C          -1.78335000453481     -0.00000008735485     -1.78334999338454\\
  C          -0.00000000105234      1.78335005429209     -1.78335005156630\\
  C           1.78335002027435     -0.00000001363369     -1.78335001720364\\
  C          -0.00000007144379     -1.78334998282724     -1.78334999157044\\
  C           1.78335002175576     -1.78335002270347     -0.00000001532119\\
  C          -0.89167504280346      2.67502505624215      0.89167512328334\\
  H           0.62931174774339      2.41266181159000      2.41266181755462\\
  C          -0.89167503841887      0.89167509366697      2.67502502119118\\
  C           2.67502507325523      0.89167512835125     -0.89167504953424\\
  H           2.41266181383066      2.41266181674156      0.62931174160635\\
  C           0.89167510897695      2.67502504312951     -0.89167503185220\\
  C           0.89167515761413     -0.89167503306512      2.67502507775177\\
  H           2.41266182946056      0.62931174594980      2.41266182557094\\
  C           2.67502510706174     -0.89167502329189      0.89167515947595\\
  C           0.89167495573222     -2.67502495181201      0.89167496165431\\
  H          -0.62931180081519     -2.41266177367791      2.41266178101083\\
  C          -2.67502499037304     -0.89167494767778     -0.89167496225488\\
  H          -2.41266179039979     -2.41266178866505      0.62931180192160\\
  C          -0.89167496407850     -2.67502500368842     -0.89167496592469\\
  H          -2.41266176945502     -0.62931179986851      2.41266178232088\\
  C          -2.67502493255006      0.89167496002748      0.89167494068849\\
  H          -2.41266177087260      2.41266177855677     -0.62931180538097\\
  C          -0.89167494056713     -0.89167496449800     -2.67502499090817\\
  H          -2.41266178444100      0.62931180868075     -2.41266178802517\\
  H          -0.62931180549674      2.41266177472925     -2.41266176512743\\
  C           0.89167496615779      0.89167493468159     -2.67502492656828\\
  H           2.41266178186247     -0.62931179801800     -2.41266177389097\\
  H           0.62931180243153     -2.41266179094420     -2.41266178804187\\
  H           2.41266177964280     -2.41266177303919     -0.62931179989037\\
  C          -0.00000002924902      3.56669992472874     -0.00000003465185\\
  H          -1.52098678825475      3.30433677854435      1.52098677343996\\
  H          -1.52098678439019      1.52098677146907      3.30433678998579\\
  C          -0.00000005171278     -0.00000002112087      3.56669996952352\\
  C           3.56669989247904     -0.00000004346810     -0.00000004653136\\
  H           3.30433677669719      1.52098677511306     -1.52098678466662\\
  H           1.52098677534344      3.30433677975171     -1.52098678972868\\
  H           1.52098676468535     -1.52098679264226      3.30433677720286\\
  H           3.30433677428362     -1.52098679072147      1.52098676796589\\
  C           0.00000001488620     -3.56670000295574      0.00000000663688\\
  H           1.52098680038067     -3.30433679171400      1.52098679732569\\
  C          -3.56670001599764      0.00000000570515      0.00000001431636\\
  H          -3.30433678761044     -1.52098679646898     -1.52098679692777\\
  H          -1.52098679507479     -3.30433678582463     -1.52098679551078\\
  H          -3.30433679365161      1.52098679972410      1.52098679993561\\
  C           0.00000000076920      0.00000001865670     -3.56670002145678\\
  H          -1.52098679793305     -1.52098679584610     -3.30433678749924\\
  H           1.52098679845831      1.52098680122413     -3.30433679394267\\
  H           0.62931178957327      4.19601179940864      0.62931178977367\\
  H          -0.62931179024357      4.19601178349398     -0.62931179532270\\
  H          -0.62931178979338     -0.62931179818211      4.19601177597025\\
  H           0.62931179731480      0.62931179399068      4.19601177988788\\
  H           4.19601180539830      0.62931178922340      0.62931179217895\\
  H           4.19601178520642     -0.62931179420254     -0.62931178960617\\
  H          -0.62931179497802     -4.19601177888772      0.62931179105690\\
  H           0.62931179066218     -4.19601178065861     -0.62931179793330\\
  H          -4.19601177709183     -0.62931179419259      0.62931178975756\\
  H          -4.19601177993386      0.62931179667068     -0.62931179305969\\
  H          -0.62931179459412      0.62931178991703     -4.19601177481524\\
  H           0.62931179704634     -0.62931179341857     -4.19601177956646\\
\\

MODEL-II\\
\\
  N           0.98825383757894      0.98825590044589      0.98825401287756\\
  C          -0.94246028351774     -0.94247002155445      0.92216479103427\\
  C          -0.94249360039997      0.92222541473550     -0.94251900106475\\
  C           0.92212781953238     -0.94245615818584     -0.94242661375008\\
  C           0.02373016898262      1.79205929038784      1.79205718629100\\
  C           1.79207903346279      1.79205796641499      0.02371381960852\\
  C           1.79207623669813      0.02368930805444      1.79207257257205\\
  C           0.00172827126903     -1.78858659629417      1.78386831952917\\
  C          -1.79366316342699     -1.79366346729035     -0.00324171883652\\
  C          -1.78856547184991      0.00175281211347      1.78385217221978\\
  C          -1.78857058615979      1.78387140527075      0.00174527337939\\
  C          -1.79365403870292     -0.00325128983397     -1.79366203948622\\
  C           0.00176564529250      1.78385459008247     -1.78856866178929\\
  C           1.78386786579976      0.00172411921634     -1.78857760020935\\
  C          -0.00323738694801     -1.79367056849785     -1.79365938563710\\
  C           1.78385218608145     -1.78858297862263      0.00174337327491\\
  C          -0.89167496936599      2.67502496308468      0.89167499364482\\
  C           0.89167524229421      2.67502499252588      2.67502480046060\\
  C          -0.89167491129224      0.89167488841428      2.67502502725193\\
  C           2.67502496969239      0.89167493660469     -0.89167498750168\\
  C           2.67502484611381      2.67502490604224      0.89167524491390\\
  C           0.89167495956384      2.67502498490829     -0.89167493675217\\
  C           0.89167496448517     -0.89167497809205      2.67502494295429\\
  C           2.67502480396992      0.89167532845421      2.67502485126800\\
  C           2.67502500753247     -0.89167498807360      0.89167494981474\\
  C           0.89167501467035     -2.67502501542973      0.89167503431216\\
  C          -0.89167495517370     -2.67502502192052      2.67502503724347\\
  C          -2.67502499619120     -0.89167504443233     -0.89167505187139\\
  C          -2.67502500689662     -2.67502497891286      0.89167499801542\\
  C          -0.89167501771621     -2.67502499311538     -0.89167502765644\\
  C          -2.67502507146966     -0.89167490079315      2.67502503496429\\
  C          -2.67502503765086      0.89167503160796      0.89167501923264\\
  C          -2.67502504127896      2.67502506045414     -0.89167492111368\\
  C          -0.89167503359332     -0.89167504102366     -2.67502499620652\\
  C          -2.67502501982695      0.89167497178708     -2.67502501034662\\
  C          -0.89167488505305      2.67502506361307     -2.67502508091182\\
  C           0.89167502095659      0.89167503736056     -2.67502502388029\\
  C           2.67502502886628     -0.89167496488644     -2.67502502202571\\
  C           0.89167500652215     -2.67502498661598     -2.67502498578671\\
  C           2.67502504437271     -2.67502504149676     -0.89167494020427\\
  C          -0.00099440756382      3.58634419091952     -0.00099864541689\\
  H          -1.52098681233486      3.30433680039591      1.52098679756212\\
  C           1.78334995293409      1.78334981703753      3.56670020494217\\
  H           0.26236315089008      3.30433680764887      3.30433684240475\\
  C           1.78334993184884      3.56670008340131      1.78334991212944\\
  H          -1.52098683016593      1.52098682227974      3.30433678470772\\
  C          -0.00098982421829     -0.00100993306203      3.58635304435189\\
  C           3.58636903467915     -0.00100404960362     -0.00100374068771\\
  H           3.30433680359659      1.52098681039483     -1.52098681162222\\
  H           3.30433684197519      3.30433682693307      0.26236315244756\\
  C           3.56670018445964      1.78334988751745      1.78334992706255\\
  H           1.52098680362419      3.30433680035406     -1.52098681919794\\
  H           1.52098680330887     -1.52098681539494      3.30433681062583\\
  H           3.30433685035899      0.26236314303085      3.30433684483659\\
  H           3.30433679922004     -1.52098681320005      1.52098680392506\\
  C          -0.00026340315186     -3.58460934031100     -0.00026561202675\\
  H           1.52098678128217     -3.30433677498689      1.52098677710577\\
  C          -1.78334997071971     -1.78335001146758      3.56669994311440\\
  H          -0.26236321351168     -3.30433677447641      3.30433677479297\\
  C          -1.78335000261754     -3.56669997745100      1.78334997251984\\
  C          -3.58461602582257     -0.00025699681866     -0.00026834525420\\
  H          -3.30433677368702     -1.52098677495653     -1.52098677585682\\
  H          -3.30433678168346     -3.30433678599234      0.26236320233733\\
  C          -3.56669994435534     -1.78335002989503      1.78334997044997\\
  H          -1.52098678416894     -3.30433677880635     -1.52098678299449\\
  H          -3.30433676573457     -0.26236322157867      3.30433677055826\\
  H          -3.30433676840224      1.52098677571386      1.52098677519490\\
  C          -1.78335001654448      3.56669991364989     -1.78334998987633\\
  H          -3.30433676806801      3.30433676699607     -0.26236321961856\\
  C          -3.56669994896721      1.78334997850109     -1.78335001557234\\
  C          -0.00026128821549     -0.00026846477795     -3.58461258556037\\
  H          -1.52098677944547     -1.52098677700807     -3.30433677529755\\
  H          -3.30433677612370      0.26236320765562     -3.30433677764137\\
  C          -1.78335002477285      1.78334997144249     -3.56669993464657\\
  H          -0.26236322632681      3.30433676335286     -3.30433676030805\\
  H           1.52098677672738      1.52098677567230     -3.30433677153370\\
  C           3.56669994569913     -1.78334997173471     -1.78334999200167\\
  H           3.30433677786894     -0.26236321084734     -3.30433677644574\\
  C           1.78334996860290     -1.78335000546612     -3.56669997749483\\
  H           0.26236320086451     -3.30433678730506     -3.30433678657961\\
  C           1.78334996122829     -3.56669996600057     -1.78335001583557\\
  H           3.30433677180689     -3.30433677124137     -0.26236321550665\\
  C           0.89167497985048      4.45837489874162      0.89167500887185\\
  C          -0.89167500669246      4.45837502760633     -0.89167502431197\\
  C           0.89167496189027      0.89167504404692      4.45837489528076\\
  H           2.41266178577648      2.41266180218777      4.19601176782112\\
  H           2.41266179254845      4.19601177706844      2.41266179530720\\
  C          -0.89167502490160     -0.89167500162960      4.45837501031015\\
  C           4.45837489278194      0.89167502103492      0.89167499012340\\
  C           4.45837500426443     -0.89167501310238     -0.89167500224182\\
  H           4.19601176781697      2.41266179299357      2.41266178981506\\
  C          -0.89167500320183     -4.45837499422450      0.89167501314086\\
  C           0.89167501627419     -4.45837498729953     -0.89167500421586\\
  H          -2.41266178884795     -2.41266178595283      4.19601179077481\\
  H          -2.41266178547959     -4.19601178755826      2.41266178812635\\
  C          -4.45837499704137     -0.89167499555508      0.89167502326207\\
  C          -4.45837499562331      0.89167502296688     -0.89167500367165\\
  H          -4.19601179022240     -2.41266178457362      2.41266178861902\\
  H          -2.41266178492142      4.19601179295177     -2.41266178688859\\
  H          -4.19601178961413      2.41266178757996     -2.41266178463739\\
  C          -0.89167499520927      0.89167502228444     -4.45837500753686\\
  C           0.89167502259689     -0.89167500184848     -4.45837498518368\\
  H          -2.41266178465514      2.41266178834449     -4.19601179051743\\
  H           4.19601179093520     -2.41266179024240     -2.41266178843377\\
  H           2.41266178868254     -2.41266178515042     -4.19601178770048\\
  H           2.41266178994862     -4.19601178875907     -2.41266178467182\\
  H           0.26236320958996      5.08768680652721      1.52098679150721\\
  H           1.52098679688377      5.08768679286767      0.26236320682572\\
  H          -0.26236320206012      5.08768677641673     -1.52098678822621\\
  H          -1.52098678764013      5.08768677597088     -0.26236320542861\\
  H           0.26236321228104      1.52098678534403      5.08768680062991\\
  H           1.52098679500812      0.26236319936593      5.08768678886442\\
  H          -1.52098678414311     -0.26236320882611      5.08768677945043\\
  H          -0.26236319997290     -1.52098679060767      5.08768677818701\\
  H           5.08768680416301      1.52098679047722      0.26236321115528\\
  H           5.08768678825145      0.26236320617013      1.52098679440427\\
  H           5.08768677845314     -1.52098679112990     -0.26236320172318\\
  H           5.08768677941746     -0.26236320656029     -1.52098679004533\\
  H          -0.26236319667222     -5.08768677677805      1.52098678900886\\
  H          -1.52098679075332     -5.08768678019881      0.26236320610929\\
  H           0.26236319950840     -5.08768678272579     -1.52098678921790\\
  H           1.52098678841722     -5.08768678102421     -0.26236320479913\\
  H          -5.08768677883633     -1.52098679153767      0.26236319685283\\
  H          -5.08768677970472     -0.26236320561875      1.52098678734833\\
  H          -5.08768677831725      1.52098678553745     -0.26236319505089\\
  H          -5.08768678104202      0.26236320429554     -1.52098679042264\\
  H          -1.52098679206809      0.26236319609186     -5.08768677541700\\
  H          -0.26236320595051      1.52098678716226     -5.08768677837020\\
  H           1.52098678615495     -0.26236319718840     -5.08768678077080\\
  H           0.26236320458641     -1.52098679134267     -5.08768678198630\\
\\

MODEL-III \\
\\
  N           0.99270851328593      0.99273651005047      0.99271946471133\\
  C          -0.94689473888554     -0.94691009685447      0.91809894625023\\
  C          -0.94692620925455      0.91814461915208     -0.94693849828178\\
  C           0.91807452574970     -0.94689534717351     -0.94687424542146\\
  C           0.04304017232318      1.78745463155201      1.78743617562678\\
  C           1.78745208090091      1.78745331698136      0.04303623963095\\
  C           1.78745567613731      0.04302152086469      1.78746945648310\\
  C          -0.01036282453086     -1.78088382166474      1.77423399430805\\
  C          -1.79204644048013     -1.79205198122798      0.00400882523144\\
  C          -1.78088085635067     -0.01036278187521      1.77422207199516\\
  C          -1.78087581886828      1.77423109423933     -0.01035263626141\\
  C          -1.79204341391868      0.00400317615897     -1.79205033805854\\
  C          -0.01035262332140      1.77423531821579     -1.78088652623629\\
  C           1.77422620710984     -0.01036556968975     -1.78087332492307\\
  C           0.00401436478744     -1.79205350948245     -1.79204380615728\\
  C           1.77422748559281     -1.78088725121058     -0.01036601854845\\
  C          -0.88002248038601      2.66614252334967      0.88644370671354\\
  C           0.90332968623953      2.69596409723028      2.69595535601302\\
  C          -0.88001914636129      0.88644360228502      2.66613321234421\\
  C           2.66612938598677      0.88644774718905     -0.88001380478200\\
  C           2.69596174540629      2.69595743537000      0.90333709415322\\
  C           0.88643968290209      2.66615295807465     -0.88002020157729\\
  C           0.88644301566782     -0.88001475040028      2.66614234868785\\
  C           2.69595445149890      0.90334357029478      2.69596124393509\\
  C           2.66613462780031     -0.88001631914245      0.88644350959157\\
  C           0.89124974306571     -2.67651457399484      0.89124035718785\\
  C          -0.89858048589705     -2.68806359147440      2.68869012752650\\
  C          -2.68091589751407     -0.89599295220164     -0.89598654724613\\
  C          -2.69284273187393     -2.69284587560966      0.89799399425035\\
  C          -0.89598915452063     -2.68092642148076     -0.89599359995085\\
  C          -2.68806986488825     -0.89859651468928      2.68869654143909\\
  C          -2.67650984242697      0.89123438544410      0.89124860433825\\
  C          -2.68806480259289      2.68869184991878     -0.89858427520762\\
  C          -0.89598718854168     -0.89598976199208     -2.68091648809581\\
  C          -2.69284792187788      0.89799807484009     -2.69284676130985\\
  C          -0.89859678270434      2.68869992292138     -2.68807803454561\\
  C           0.89124157242926      0.89124483156493     -2.67651067572592\\
  C           2.68868623977628     -0.89858283869810     -2.68805822903998\\
  C           0.89798842357864     -2.69283984242944     -2.69284209712480\\
  C           2.68869689743287     -2.68807297921111     -0.89859524249467\\
  C          -0.00039109433174      3.57249331327827     -0.00039027142474\\
  C          -1.78796000645336      3.57104916183916      1.79283626295091\\
  C           1.78335022710354      1.78334992480106      3.56669990959977\\
  C           0.00000022327672      3.56669996231089      3.56669998539962\\
  C           1.78335000990749      3.56670006055099      1.78335006967611\\
  C          -1.78795426050965      1.79283343135625      3.57105052216168\\
  C          -0.00039454370754     -0.00038726667909      3.57249035172839\\
  C           3.57249439755139     -0.00039064531674     -0.00039254878895\\
  C           3.57104814429656      1.79284218750208     -1.78796059194987\\
  C           3.56669998476201      3.56669965516019      0.00000024614985\\
  C           3.56669986520142      1.78335010904662      1.78335009148931\\
  C           1.79283462112595      3.57105743563248     -1.78796061878791\\
  C           1.79283969945080     -1.78796228407617      3.57105429779967\\
  C           3.56669983694584      0.00000046957115      3.56670010345875\\
  C           3.57105860336099     -1.78796295853867      1.79284138115589\\
  C          -0.00126137276629     -3.57408683934805     -0.00125951116582\\
  C           1.79191858487776     -3.57601696123521      1.79191993658259\\
  C          -1.78335000521494     -1.78334990097057      3.56669999365722\\
  C          -0.00000017090749     -3.56670016220645      3.56670020936343\\
  C          -1.78334997696250     -3.56670003426781      1.78334990210399\\
  C          -3.57408804451145     -0.00126023016125     -0.00126395900010\\
  C          -3.57841119826352     -1.79355750206405     -1.79355801037086\\
  C          -3.56669999779069     -3.56669988918388      0.00000013847412\\
  C          -3.56669999478561     -1.78334999462831      1.78335003923789\\
  C          -1.79355421380840     -3.57841496720667     -1.79355608409718\\
  C          -3.56669997048685     -0.00000010386853      3.56669998313404\\
  C          -3.57601688422027      1.79192008326171      1.79191747481915\\
  C          -1.78334991776357      3.56669995638166     -1.78334998884780\\
  C          -3.56670015525610      3.56670022346141     -0.00000017856438\\
  C          -3.56670001867753      1.78334992589604     -1.78334998835397\\
  C          -0.00125935867429     -0.00126513183360     -3.57408638833723\\
  C          -1.79355558965973     -1.79355628515405     -3.57841002701361\\
  C          -3.56670008816359      0.00000022677319     -3.56670005211290\\
  C          -1.78334994507673      1.78334994819071     -3.56669994737875\\
  C          -0.00000011805422      3.56670008904033     -3.56670000021559\\
  C           1.79191953489782      1.79191823753114     -3.57601663350195\\
  C           3.56669994422731     -1.78335000562314     -1.78334999304827\\
  C           3.56670023009359     -0.00000017845886     -3.56670013030529\\
  C           1.78334996180161     -1.78334998611945     -3.56670002687692\\
  C           0.00000017365845     -3.56670005159827     -3.56670002378548\\
  C           1.78335000464300     -3.56669994777210     -1.78334994772445\\
  C           3.56670012620605     -3.56670009098298     -0.00000008463812\\
  C           0.90136811641756      4.46963510667402      0.90136519731206\\
  C          -0.90193019430320      4.46903677675829     -0.90193090840222\\
  C          -2.67502477193961      2.67502480936306      2.67502481107906\\
  C          -2.67502485483698      4.45837480217333      0.89167507916032\\
  C          -0.89167510572408      4.45837494660587      2.67502489923495\\
  C           0.90136699541907      0.90137280448212      4.46963447717928\\
  H           2.41266173014619      2.41266177440232      4.19601177449335\\
  H           0.62931173938227      4.19601178880982      4.19601179157615\\
  C          -0.89167502477955      2.67502494994186      4.45837498035391\\
  H           2.41266176055674      4.19601174834980      2.41266174714754\\
  C          -2.67502490139865      0.89167504408157      4.45837497072425\\
  C          -0.90192946673921     -0.90193041760333      4.46903361812632\\
  C           4.46962813012423      0.90136744853182      0.90137019388487\\
  C           4.46903012640448     -0.90192826958563     -0.90192704892254\\
  C           2.67502482597235      2.67502481144408     -2.67502479251607\\
  C           4.45837475849720      0.89167506194279     -2.67502486090155\\
  C           4.45837497692887      2.67502502606536     -0.89167502233503\\
  H           4.19601178572576      4.19601187878402      0.62931173291186\\
  C           2.67502499368282      4.45837511099320     -0.89167505857380\\
  H           4.19601179626664      2.41266174510452      2.41266176045711\\
  C           0.89167500283504      4.45837483259074     -2.67502489471086\\
  C           2.67502480845190     -2.67502477556423      2.67502482299042\\
  C           0.89167501347816     -2.67502480201241      4.45837477747435\\
  C           2.67502497086966     -0.89167520651871      4.45837480056638\\
  H           4.19601182322939      0.62931169778906      4.19601174777201\\
  C           4.45837504499528     -0.89167510873676      2.67502496768052\\
  C           4.45837480182928     -2.67502483710291      0.89167497070095\\
  C          -0.90145015775815     -4.46986784187206      0.90146083089988\\
  C           0.90145810064309     -4.46987005526987     -0.90145163206118\\
  C           2.67502490929096     -4.45837493930108      0.89167503699271\\
  C           0.89167506598722     -4.45837491491479      2.67502484600765\\
  H          -2.41266178465720     -2.41266180479668      4.19601178504384\\
  H          -0.62931175900523     -4.19601175372126      4.19601174509313\\
  H          -2.41266177552005     -4.19601177418712      2.41266179598641\\
  C          -4.46986461633563     -0.90145185642177      0.90145905180849\\
  C          -4.46986574329488      0.90145859561279     -0.90145056497331\\
  C          -2.67502489833300     -2.67502489545591     -2.67502491948403\\
  C          -4.45837491377906     -0.89167517019673     -2.67502494876474\\
  C          -4.45837499538773     -2.67502497863310     -0.89167506888897\\
  H          -4.19601178196236     -4.19601181160593      0.62931176706547\\
  C          -2.67502497941187     -4.45837503167439     -0.89167507349512\\
  H          -4.19601178572455     -2.41266178508941      2.41266177687435\\
  C          -0.89167509133542     -4.45837494114043     -2.67502495470737\\
  H          -4.19601179896904     -0.62931177959967      4.19601178406183\\
  C          -4.45837506733931      0.89167504640461      2.67502497119300\\
  C          -4.45837490047435      2.67502484527968      0.89167506056839\\
  H          -2.41266180673945      4.19601179183122     -2.41266178347111\\
  H          -4.19601174921691      4.19601174216630     -0.62931176190568\\
  H          -4.19601177032090      2.41266178559287     -2.41266177311487\\
  C          -0.90144964570160      0.90145809060342     -4.46986746617280\\
  C           0.90145930896026     -0.90145203281637     -4.46986578301589\\
  C          -0.89167510504459     -2.67502493289373     -4.45837495185761\\
  C          -2.67502492501334     -0.89167511539020     -4.45837494734037\\
  H          -4.19601176207222      0.62931176004212     -4.19601176712994\\
  H          -2.41266179903275      2.41266179507407     -4.19601179440200\\
  H          -0.62931177155483      4.19601176926452     -4.19601179206803\\
  C           0.89167505274193      2.67502495308507     -4.45837504421133\\
  C           2.67502483544623      0.89167505396812     -4.45837489614936\\
  H           4.19601178304662     -2.41266177855048     -2.41266178384324\\
  H           4.19601174205587     -0.62931176580711     -4.19601175544173\\
  H           2.41266177763818     -2.41266177050257     -4.19601176732947\\
  H           0.62931176755901     -4.19601177145464     -4.19601177210057\\
  H           2.41266178164310     -4.19601179034383     -2.41266179253294\\
  H           4.19601176418092     -4.19601177423014     -0.62931178294445\\
  C           0.00000002240717      5.35004996424226      1.78335009174119\\
  C           1.78335005068972      5.35004986937983     -0.00000005861075\\
  C           0.00000004571617      5.35005004616322     -1.78335005219870\\
  C          -1.78335007754643      5.35005003888394     -0.00000000001429\\
  H          -3.30433688680385      3.30433687117428      3.30433687423712\\
  H          -3.30433682883144      5.08768682824537      1.52098679374228\\
  H          -1.52098680974217      5.08768682677467      3.30433681493602\\
  C          -0.00000007511104      1.78335005820701      5.35004992745559\\
  C           1.78335010118795      0.00000001642383      5.35005009462587\\
  H          -1.52098682122543      3.30433681625308      5.08768681540160\\
  H          -3.30433682284637      1.52098679210703      5.08768680716892\\
  C          -1.78335005505893     -0.00000000074413      5.35004995162944\\
  C           0.00000006284256     -1.78335012830263      5.35005007148012\\
  C           5.35005000003074      1.78335006186190     -0.00000005767018\\
  C           5.35004997491662     -0.00000006751107      1.78335006594199\\
  C           5.35005008127249     -1.78335007474903      0.00000004081907\\
  C           5.35005007072661      0.00000000596525     -1.78335008311398\\
  H           3.30433687131874      3.30433687152512     -3.30433688232229\\
  H           5.08768683446117      1.52098679939159     -3.30433682908007\\
  H           5.08768682098856      3.30433679867002     -1.52098682004574\\
  H           3.30433680037024      5.08768681406644     -1.52098681075761\\
  H           1.52098680264413      5.08768683066055     -3.30433682472405\\
  H           3.30433687638475     -3.30433688657104      3.30433687709803\\
  H           1.52098680658716     -3.30433683406627      5.08768684303497\\
  H           3.30433680360678     -1.52098679650252      5.08768684436965\\
  H           5.08768681247561     -1.52098680752532      3.30433680110339\\
  H           5.08768683543838     -3.30433683535950      1.52098680872574\\
  C          -0.00000001993767     -5.35004997907903      1.78335007872329\\
  C          -1.78334999069771     -5.35004992878763      0.00000003928588\\
  C           0.00000004609816     -5.35005000215043     -1.78335000714422\\
  C           1.78335002533870     -5.35005000076957     -0.00000002229125\\
  H           3.30433680901922     -5.08768680439140      1.52098678912045\\
  H           1.52098679561890     -5.08768680781859      3.30433681950302\\
  C          -5.35004997883142     -1.78335001618541      0.00000004350229\\
  C          -5.35004992166553      0.00000000318479      1.78335001049092\\
  C          -5.35005000384755      1.78335006170341     -0.00000002219316\\
  C          -5.35005002548781      0.00000010117243     -1.78335002453893\\
  H          -3.30433682390703     -3.30433682379177     -3.30433682388100\\
  H          -5.08768679288667     -1.52098677644695     -3.30433679034956\\
  H          -5.08768678648094     -3.30433678838343     -1.52098679252622\\
  H          -3.30433679233125     -5.08768678624231     -1.52098679113763\\
  H          -1.52098678677443     -5.08768679470808     -3.30433679665964\\
  H          -5.08768677989167      1.52098679356541      3.30433679578528\\
  H          -5.08768680480743      3.30433681838986      1.52098678545505\\
  C          -1.78335003636139      0.00000006481756     -5.35005000858268\\
  C          -0.00000000097835      1.78335002468337     -5.35004994024191\\
  C           1.78335005680019     -0.00000001362714     -5.35004999555408\\
  C           0.00000006267745     -1.78335001322693     -5.35004999820120\\
  H          -1.52098678442916     -3.30433679698560     -5.08768679555018\\
  H          -3.30433679616602     -1.52098679163340     -5.08768679351619\\
  H           1.52098678796890      3.30433679588038     -5.08768678619765\\
  H           3.30433681969031      1.52098679231611     -5.08768680731671\\
  C          -0.89167498411241      6.24172500226326      0.89167492980350\\
  H           0.62931179278082      5.97936178906575      2.41266177277019\\
  H           2.41266177695128      5.97936181677670      0.62931179968094\\
  C           0.89167492795971      6.24172502183672     -0.89167494461482\\
  H          -0.62931180265926      5.97936177731450     -2.41266178980250\\
  H          -2.41266178167101      5.97936178008288     -0.62931180559627\\
  H           0.62931180153720      2.41266177909775      5.97936180102370\\
  C          -0.89167497454016      0.89167499048282      6.24172500767039\\
  C           0.89167488761157     -0.89167491705367      6.24172493371404\\
  H           2.41266177107952      0.62931179444970      5.97936177429739\\
  H          -2.41266178142456     -0.62931180721663      5.97936178711919\\
  H          -0.62931180423318     -2.41266178231216      5.97936177986797\\
  C           6.24172495470822      0.89167493714839     -0.89167495930678\\
  H           5.97936178466338      2.41266178003118      0.62931179086097\\
  H           5.97936179349117      0.62931180508265      2.41266177677207\\
  C           6.24172495507967     -0.89167493162609      0.89167494212567\\
  H           5.97936177047261     -2.41266178241711     -0.62931180158811\\
  H           5.97936177505975     -0.62931180599430     -2.41266177489811\\
  C           0.89167499608763     -6.24172500370621      0.89167499614140\\
  H          -0.62931178844012     -5.97936178556458      2.41266177272830\\
  H          -2.41266178379577     -5.97936179667691      0.62931179315198\\
  C          -0.89167503435531     -6.24172500772639     -0.89167501649230\\
  H           0.62931179133686     -5.97936177442052     -2.41266178623756\\
  H           2.41266177921279     -5.97936177527807     -0.62931179692988\\
  C          -6.24172497975938     -0.89167503344404     -0.89167502646753\\
  H          -5.97936177806575     -2.41266177736764      0.62931178353862\\
  H          -5.97936179290187     -0.62931179836572      2.41266178332785\\
  C          -6.24172500929417      0.89167498637189      0.89167500321771\\
  H          -5.97936177177583      2.41266177796808     -0.62931179389239\\
  H          -5.97936176706148      0.62931179078594     -2.41266177208376\\
  C          -0.89167501149649     -0.89167503624769     -6.24172497703068\\
  H          -2.41266177951478      0.62931177982780     -5.97936177509951\\
  H          -0.62931179925213      2.41266177899083     -5.97936179008787\\
  C           0.89167499705405      0.89167498784013     -6.24172501268322\\
  H           2.41266178397666     -0.62931179492606     -5.97936177333529\\
  H           0.62931179155237     -2.41266177761767     -5.97936177340809\\
  C           0.00000000937927      7.13339995361926     -0.00000000014701\\
  H          -1.52098679228189      6.87103677739635      1.52098679593650\\
  H           1.52098679840088      6.87103677207053     -1.52098680140280\\
  H          -1.52098678605511      1.52098678413908      6.87103678312772\\
  C          -0.00000000494873     -0.00000001498379      7.13339998066502\\
  H           1.52098679968020     -1.52098680172144      6.87103677709242\\
  C           7.13339999777921     -0.00000001543246     -0.00000000911847\\
  H           6.87103678103422      1.52098679557714     -1.52098679562810\\
  H           6.87103677709517     -1.52098680072143      1.52098679807651\\
  C           0.00000001419405     -7.13339996165242     -0.00000000602489\\
  H           1.52098679173145     -6.87103677077507      1.52098678780718\\
  H          -1.52098678704338     -6.87103677039356     -1.52098678721355\\
  C          -7.13339996297749      0.00000000419993     -0.00000000206184\\
  H          -6.87103677650425     -1.52098678160630     -1.52098678673803\\
  H          -6.87103676776098      1.52098679261586      1.52098679130147\\
  C          -0.00000000737002      0.00000000950402     -7.13339996593978\\
  H          -1.52098678207390     -1.52098678694495     -6.87103677668752\\
  H           1.52098679239184      1.52098679139079     -6.87103676840001\\
  H           0.62931179294585      7.76271177635496      0.62931179106485\\
  H          -0.62931179650477      7.76271177782203     -0.62931179598953\\
  H          -0.62931178956894     -0.62931179474378      7.76271177488492\\
  H           0.62931179542534      0.62931179291469      7.76271177438385\\
  H           7.76271176603003      0.62931179726270      0.62931179910892\\
  H           7.76271177239448     -0.62931179544806     -0.62931179086020\\
  H          -0.62931179150361     -7.76271178109359      0.62931179090930\\
  H           0.62931179278223     -7.76271177477175     -0.62931179726875\\
  H          -7.76271178024067     -0.62931179054060      0.62931179288899\\
  H          -7.76271177418562      0.62931179761433     -0.62931179146035\\
  H          -0.62931179045533      0.62931179233087     -7.76271177585393\\
  H           0.62931179751800     -0.62931179173578     -7.76271177438461\\
\\

\newpage

\section{Computational details of CRPA+FCI calculations}
Here we use a minimal basis, consisting of 4 spatial orbitals (8 spin orbitals). These were obtained by wannierization and projection of the localized Bloch states on $sp^3$ orbitals on the nearest neighbour atoms. In Tab.~\ref{tab:crpafcienergies}, the state energies and corresponding spin are tabulated. 
\begin{table}[H]
\centering 
    \caption{Energies and spin from CRPA+FCI calculations.}
    \label{tab:crpafcienergies}
\begin{tabular}{cc}
    Energy wrt gs (eV) & $S$ \\ \hline 
    0.00 & 1 \\
    0.49 & 0 \\ 
    0.50 & 0  \\
    1.38 & 0 \\
    2.19 & 1 \\
    2.20 & 1 \\
    3.18 & 0 \\
    3.20 & 0\\
    4.54 & 1 \\
    4.55 & 1 \\
    4.99 & 0 \\
    5.13 & 0 \\ 
    5.15 & 0\\ \hline
\end{tabular}
    \caption{Energies and spin from CRPA+FCI calculations.}
    \label{tab:crpafcienergies}
\end{table}

\newpage

\section{Supplemetary discussions}

\subsection{Comparability of experimental zero-phonon lines and computed vertical energies}

Even though the vertical energies are, in a strict sense, not to be directly compared to absorption ZPLs - as geometry relaxation effects and vibrational energy corrections are neglected -, both aforementioned factors are expected to be low in a soid-phase environment. For example, the distance of the ZPL line from the maximum-intensity peak of the phonon side band (which approximately equals to the difference between the ZPL and the vertical gap) is 0.23 eV for $^3A_2 \rightarrow ^3E$) and only 0.13 eV for $^3E' \rightarrow ^3E''$ (see SI of Ref.~\cite{Luu2024}).

Furthermore, we can expect the error cancellation of geometry relaxation effects if the $^3A_2$ state is not involved in the transition of interest. In these cases, CASSCF-NEVPT2 reproduces the TA peak positions with an error margin of less than 0.2 eV. On the other hand, the energy gap of $^3A_2 \rightarrow ^3E$ is overestimated by more than 0.3 eV. This anomaly is not surprising, as the geometry of $^3A_2$ was used for the computations.   

\subsection{Derivation of oscillator strengths at CASSCF-NEVPT2 level}
\label{si:sect:osc}
The oscillator strength ($f_{\rm osc}$) between the initial (i) and final (f) state of absorption can be calculated as

\begin{equation}
f_{{\rm osc}, i \rightarrow f}=\frac{2m_e}{3e^2\hbar^2}(E_f - E_i) \mu_{i \rightarrow f}^2    
\label{eq:fosc}
\end{equation}

According to Eq.~\ref{eq:fosc}, the cross section of a given absorption process (manifested as area under the respective TA peak in the measurement), is determined by the energy difference ($E_f - E_i$) and the electric transition dipole monent ($\mu_{i \rightarrow f}$). The constants of $m_e$, $e$ and $\hbar$ refer to the mass of electron, the elementary charge, and to the reduced Planck constant, respectively.

Unfortunately, the present common implementations of the NEVPT2 method only account for correction of the energy but not of the wave function, hence the transition dipole moment. Accordingly, we obtain $\mu_{i \rightarrow f}$ at CASSCF level, which, however, is expected to be reasonable for localized band-gap states. Nevertheless, as the sensitivity of $f_{\rm osc}$ to the $\mu_{i \rightarrow f}$ value is large ($\mu_{i \rightarrow f}^2$ appears in Eq.~\ref{eq:fosc}), we can only expect the determination of the order of magnitude of $f_{\rm osc}$. 

\subsection{Discussion of the oscillator strength of the $^1A_1 \rightarrow ^1E''$ transition}

In the following, we gather the possible reasons of the outstandingly large deviation between the theoretical and experimental $f_{\rm osc}$ of the $^1A_1 \rightarrow ^1E''$ absorption process.

(i) First, the large contribution of NEVPT2 to the transition energy of $^1A_1 \rightarrow ^1E''$ (almost 5 eV decrease relative to CASSCF, see SI, Tab.~\ref{tab:singlet_spectrum}) might indicate that valence-band and conduction-band orbitals have a significant influence on the wavefunction. It is also conspicuous that the $^1E"$ shows the slowest convergence of CASSCF and NEVPT2 energy contributions with respect to the model size. Therefore, in this case, the transition matrix element, computed at CASSCF level of theory without NEVPT2 correction (see SI section~\ref{si:sect:osc}), might be a considerable source of error in $f_{\rm osc}$. (ii) Alternatively, our assignment of the peak to the $^1E''$ state might be incorrect, albeit we note that all possible configurations were investigated in this work considering the four band-gap orbitals. Here, it is worth mentioning that excitations to $e'$ orbitals (the analogues of $e$ orbitals in the 3rd, unoccupied electronic shell) were previously suggested to form additional band-gap states.\cite{Bhandari2021} We could not confirm these results by extending the active space, as we found that such $e'$ orbitals are deeply embedded in the conduction band, lying more than 10 eV above the semi-occupied $e$ orbitals. (See SI, section S1.6.). 
(iii) Finally, we also do not exclude the possibility that a hitherto unidentified factor in the experimental design causes an unexpectred drop in the TA peak intensity of the $^1A_1 \rightarrow ^1E''$ excitation.   

\section*{References}
\vskip 0.5cm